\documentclass[aps,pre,twocolumn,superscriptaddress]{revtex4-1}

\usepackage{epsfig}
\usepackage{graphicx}
\usepackage{dcolumn}
\usepackage{bm}
\usepackage{amsmath,amssymb}
\usepackage{booktabs}
\usepackage{threeparttable}
\usepackage{subfigure,dcolumn}
\usepackage{threeparttable}
\usepackage{url}

\begin{document}

\title{Learning epidemic threshold in complex networks by Convolutional Neural
Network}

\author{Qi Ni}
\affiliation{School of Communication and Electronic Engineering, East China Normal University, Shanghai 200241, China}

\author{Jie Kang}
\affiliation{School of Communication and Electronic Engineering, East China Normal University, Shanghai 200241, China}

\author{Ming Tang}\email{tangminghan007@gmail.com}
\affiliation{School of Mathematical Sciences, Shanghai Key Laboratory of PMMP, East China Normal University, Shanghai 200241, China}
\affiliation{Shanghai Key Laboratory of Multidimensional Information Processing, East China Normal University, Shanghai 200241, China}

\author{Ying Liu}\thanks{shinningliu@163.com}
\affiliation{School of Computer Science, Southwest Petroleum University, Chengdu 610500, China}
\affiliation{Big Data Research center, University of Electronic Science and Technology of China, Chengdu 610054, China}

\author{Yong Zou}
\affiliation{School of Physics and Electronic Science, East China Normal University, Shanghai 200241, China}

\date{\today}

\begin{abstract}

Deep learning has taken part in the competition since not long ago to learn and identify
phase transitions in physical systems such as many body quantum systems,
whose underlying lattice structures are generally regular as they're
in euclidean space. Real networks have complex structural features
which play a significant role in dynamics in them, and thus the structural
and dynamical information of complex networks
can not be directly learned by existing neural network models.
Here we propose a novel and effective framework to
learn the epidemic threshold in complex networks by combining the structural and
dynamical information into the learning procedure.
Considering the strong performance of learning in Euclidean space,
Convolutional Neural Network (CNN) is used and,
with the help of \textit{confusion scheme},
we can identify precisely the outbreak threshold of epidemic dynamics.
To represent the high dimensional network data set in Euclidean space for CNN,
we reduce the dimensionality of a network by using graph representation learning algorithms
and discretize the embedded space to convert it into an image-like structure. We then
creatively merge the nodal dynamical states with the structural embedding by
multi-channel images. In this manner, the proposed model can draw the conclusion
from both structural and dynamical information.
A large number of simulations show a great performance
in both synthetic and empirical network data set.
Our end-to-end machine learning framework is robust
and universally applicable to complex networks with arbitrary
size and topology.

\end{abstract}

\maketitle

\textbf{The combination of machine learning and complex networks isn't a new topic
since a variety of graph representation learning techniques and deep learning based
graph neural networks have found their applications in the fields of network embedding,
node classification, link prediction and network reconstruction, etc. However, these
research frontiers focus on learning the topology instead of dynamics of the networks.
Deep learning algorithms have been applied in learning and
identifying the dynamics in Ising model these years. Inspired by previous researches, we propose
in this work a robust and efficient machine learning framework to identify the
epidemical threshold of SIS dynamics in complex networks. Convolutional Neural Network
(CNN) plays a key role of learner in the framework and the input networks are converted
into 2D image-like structures, on which CNN performs well. We then creatively
add dynamical information into the images and thus our model can learn the networks
by gathering the structural and dynamical information together. A large number of
experiments on both synthetic and empirical networks prove that
the proposed framework is accurate and universally applicable.
The proposition of our framework opened a new perspective to learn
and identify critical point of dynamics in complex networks, using machine
learning techniques. Our work combines both topological and dynamical
information of a network, exploring a new way to solve the threshold-related issues.
}

\section{Introduction} \label{sec:intro}

Deep learning is an interdisciplinary frontier as researches on computer vision,
natural language processing, recommendation system, etc., thrive~\cite{JM:2015}.
This benefits much from the increasing availability of computational resources and
easiness to access massive data set.
Recently, this topic has attracted much attention from
physicists because of the strong performance of detecting, predicting and uncovering various
phases of matter in quantum many-body systems~\cite{Wang:2016,OO:2016,SRN:2017,
ZMK:2017,vNLH:2017,CM:2017,CT:2017,ZK:2017,LQMF:2017,DLD:2017a,DLD:2017b,VKK:2018}.
Neural network based models play an irreplacable role in this field because they're
able to not only learn generate phases or states of matter that are previously known
~\cite{CM:2017,DLD:2017b,ZMK:2017,SRN:2017} or uncovering phase transitions
~\cite{Wang:2016,OO:2016,vNLH:2017}, but predict out-of-equilibrium phases of matter
that have not been known yet~\cite{VKK:2018}. Both Feed-Forward Neural Network (FFNN)
and Convolutional Neural Network (CNN) have been exploited to learn and discover the
hidden pattern of phase transition in Ising-type lattice structures~\cite{CM:2017}.
Although FFNN can give a good prediction sometimes, it is usually outperformed by CNN
because the latter is able to store and learn the topological information of the system. It
was demonstrated that the threshold or the critical point phase transition in the
Ising model can be predicted through deep learning~\cite{DL:book}. All these
accomplishments benefit greatly from the Euclidean topology of the underlying Ising
spin lattice.

In recent years, researches have achieved unprecedented results
on a large variety of natural, social, and
engineering systems~\cite{Newman:book}. A complex network, by definition, has a
complex topology and, as a natural phenomenon associated with network dynamics,
the emergence of distinct phases and phase transitions are ubiquitous
~\cite{DGM:2008}. Though there have been a massive number of works
incorporating machine learning into the field of complex networks
~\cite{PAS:2014,GL:2016,KW:2016,WCZ:2016,HYL:2017}, these studies were confined to
learning the {\em structural information} of the network. Network
representation learning is a good example in which a network is dimensionally
reduced with most of its structure information intact. There are applications
in problems such as link prediction~\cite{LK:2007}, clustering~\cite{DHZGS:2001}, and
node classification~\cite{BCM:2011}. Based on random walk and graph search algorithm,
the algorithm named ``node2vec'' can embed the network topology into a lower
dimensional space by integrating macroscopic and microscopic structural information
of the network~\cite{GL:2016}. The algorithm ``\textit{Deepwalk}'' finds a way to unite
skipgram, a natural language processing technique, with random walk sequences
on graphs~\cite{PAS:2014}, providing a concise solution to graph embedding.
Furthermore, it was proved that the performances of deep learning models
such as Graph Convolutional Networks (GCN)~\cite{KW:2016} and structural
deep network embedding~\cite{WCZ:2016} can be equivalent to those of
traditional, random-walk based methods.

The key aspect that distinguishes
our work from the previous works is that we exploit machine learning
to deal with {\em dynamics} responsible for phase transitions in complex
networks.
To be concrete, we exploit deep learning, together with \textit{confusion scheme}
~\cite{vNLH:2017} to identify phase transitions of epidemic spreading in complex
networks~\cite{PSV:2001a,PSV:2001b,Newman:2002,Vespignani:2009,
Vespignani:2012,PSCVV:2015,DGPA:2016,WTSB:2017}.
There is a typical second-order phase transition in this dynamical process.
It is an active research topic to precisely learn and identify
the critical point of the phase transition.
A widely applied method is
the so-called ``degree-based mean field'' approach~\cite{PSV:2001a,DGM:2008},
which gives the theoretical threshold as
$\left\langle k\right\rangle/{\left\langle k^2\right\rangle}$,
where $\left\langle k\right\rangle$ and $\left\langle k^2 \right\rangle$ are
the first and second moments of the degree distribution, respectively.
However, the theoretical prediction is not always accurate because of the
dynamical correlations among nodes.
Thus for real world networks, there must be a discrepancy between the
calculated and simulation results.
It is commonly agreed that the results obtained by simulation are accurate enough
by using Monte-Carlo simulations through measures such as
network susceptibility~\cite{FCPS:2012}, variability~\cite{SWTD:2015} or the
average lifetime~\cite{BCPS:2013}. The basic idea and working principle of
our machine learning based approach differ fundamentally from the
existing methods.

We've made a significant step forward to make the deep learning model learn
and identify the critical point of phase transition~\cite{mypaper1}. However,
FFNN fails to take the network topology into consideration and the
structural information is omitted at the
very beginning in the input layer. To overcome the disadvantage, a couple of sampling
methods based on nodal importance are proposed to manually ``restore'' the missing
topological information. Inspired by our previous work, we use CNN to
learn both topological and dynamical information in networks and make the
prediction all by itself without any additional operation and feature engineering.
Also, our CNN based model outperforms our previous model and no
sampling method is needed to get topological information.

The rest of the work is organized as follows: in section~\ref{sec:definition} we focus on
the susceptible-infected-susceptible (SIS) epidemic spreading model and the structure \
and workflow of the proposed machine learning framework.
In section ~\ref{sec:hyperbolic}, we
introduce hyperbolic embedding in complex networks and explain in detail why
it is needed in our work, together with the experimental results in synthetic networks.
In section~\ref{sec:real} we show the results and
performances of the proposed model in real-world networks.
Finally in section~\ref{sec:conclusion}, we take a review of our contribution
in this work and provide a couple of open questions to inspire further researches.

\section{Proposed model and method} \label{sec:definition}

In this section we introduce the SIS spreading
model, as well as the proposed CNN based machine learning framework and its training
data set.

\subsection{Epidemic spreading and measure of susceptibility}

In this work, the classical SIS epidemic process~\cite{Newman:book} is used. At each
time step, an infected node transmits the disease to its susceptible neighbors
with infection rate $\beta$, and the infected node returns to susceptible state
with recovery rate $\mu$. As we know, there are two phases in the
spreading process: active and absorbing state. In active state there are both two
types of nodes in the network while in absorbing state, there is no infected node.
The effective infection rate is defined as the ratio of the infection rate to the
recovery rate, which is  $\lambda=\beta/\mu$, and the critical point of $\lambda$
can be denoted as $\lambda_c$. The system approaches the absorbing state after a
period of time if $\lambda\leq\lambda_c$. For $\lambda>\lambda_c$, the system will
reach an endemic state where the density of the infected nodes remains stable.
Under this circumstance, the system is in the active phase.

The numerical threshold of the SIS process can be characterized
by susceptibility~\cite{FCPS:2012} defined as
\begin{equation} \label{susceptibility}
\chi=N\frac{\left\langle \rho^2\right\rangle-{\left\langle \rho\right\rangle}^2}{\left\langle \rho\right\rangle},
\end{equation}
where $\rho$ is the density of the infected nodes (i.e., the order parameter),
$N$ is the network size, $\left\langle\rho\right\rangle$ and
$\left\langle\rho\right\rangle^2$ are the first and second moments of $\rho$,
respectively. The susceptibility measure indicates the variation in outbreak sizes
over many runs with the same parameter $\lambda$.
The order parameter associated with the second-order phase transition ($\rho$ in this case)
shows a power-law distribution near the critical point. Hence the variation of
the order parameter reaches the peak at this point and so does the susceptibility
curve. As a function of the effective infection rate $\lambda$, the
susceptibility measure reaches its maximum value at $\lambda_c^\chi$, which is the
threshold of the epidemic process or the critical point of phase transition.
In machine learning, especially for supervised learning, what we need is a
measure to accurately reflect the ground truth of the transition point and the
corresponding label during training can be obtained. The susceptibility measure plays an
important role under this circumstance.

\subsection{Proposed machine learning framework: overview}

\begin{figure*}
\centering
\epsfig{file=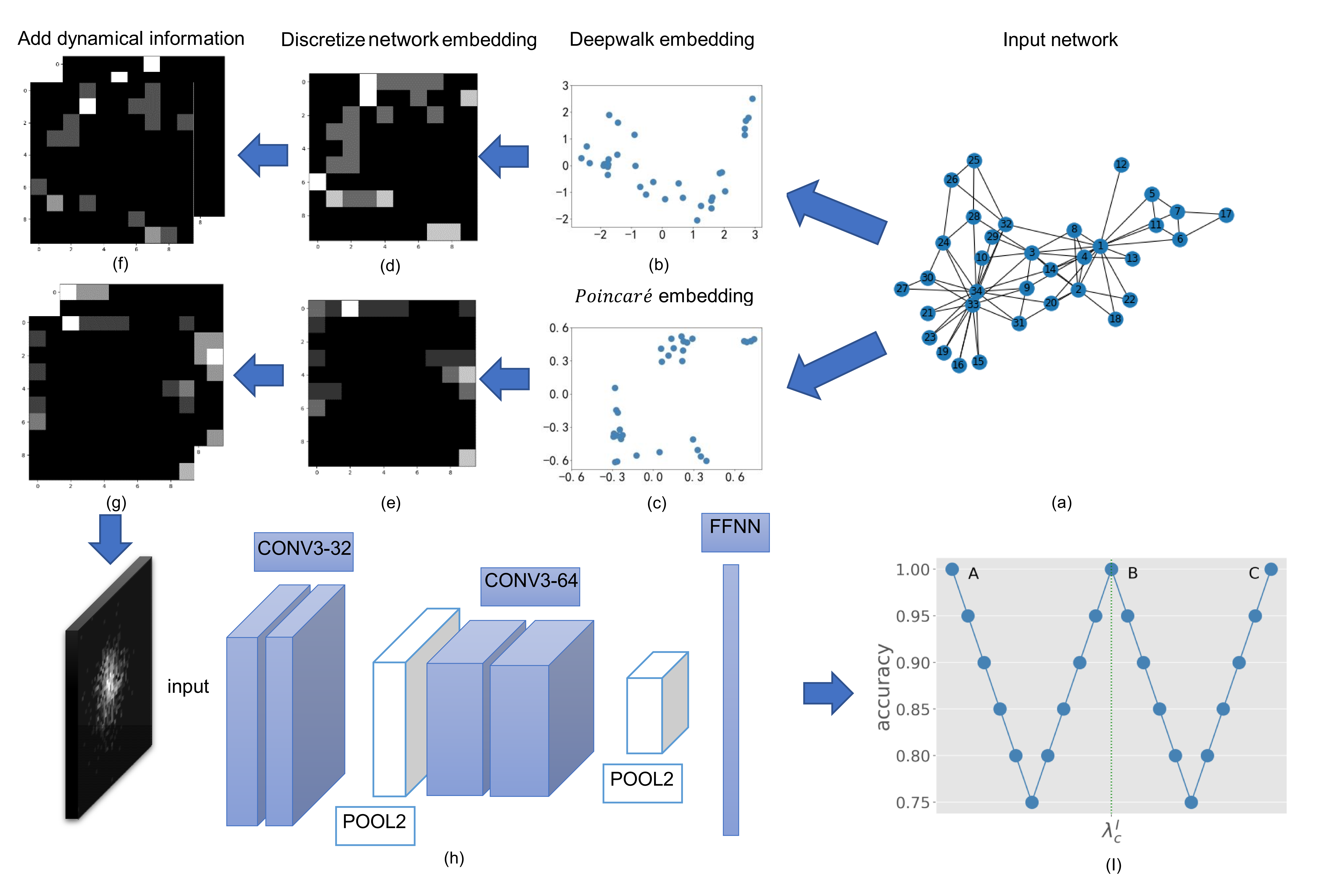,width=1\linewidth}
\caption{
\textit{Workflow of the proposed machine learning model.}
(a) Input network, which is represented by adjacency matrix sized $N*N$, with the
$N$ denoting the number of nodes.
(b, c) Network representation learning. Network dimensional reduction algorithms and
PCA are used to reduce the dimensionality of the input signal to 2.
(d, e) Discretize the continuous embedded data into histograms, as this image-like
structure makes the following learning procedure possible. The color of each
rectangle sub-region represents the number of nodes falling into it.
(f, g) Adding dynamical information. We combine the structural and dynamical
information by split the single-channel images into two channels, representing the
nodes in the state of $S$ and $I$, respectively.
(h) The architecture of CNN we use. It consists of two convolution layers with $32$
and $64$ filters, respectively. Following every convolution layer is a $(2,2)$ max
pooling layer. The second convolution-pooling block sends its output to a
$128$-neuron dense layer to calculate the probability that the input data is
supposed to be in a specific phase. All of these above are constructed with
$tensorflow$.
(I) Schematic illustration of confusion scheme to detect phase
transition and identify the epidemical threshold. For a range
of tentatively assigned threshold values, the curve of classification
accuracy of supervised learning versus the threshold value will exhibit a
$W$-shape, if there is a phase transition associated with epidemic dynamics
on the network. The location of the middle peak gives an accurate prediction
of the true threshold value. If the network dynamics do not exhibit a phase
transition, the curve would exhibit a $U$ shape (see text for a detailed
reasoning).
}
\label{fig:framework}
\end{figure*}

We propose a machine learning framework to learn the SIS dynamics and
identify the critical point of the phase transition. We turn a network
into an image-like regular structure with nodal dynamical information added by
multiple channels. The CNN can take these image-like data as input and finally
output a probability indicating how possible the input data is in one specific
phase. These outputs are compared with man-made fake labels and the critical
point can thus be inferred by \textit{confusion scheme}~\cite{vNLH:2017}, a powerful
tool combining ideas from
both supervised and unsupervised learning. For simplicity, the workflow can be
decomposed into four steps as illustrated in Fig.~\ref{fig:framework}:

(1) Embed network into two dimensional space.

(2) Discretize network embedding and add dynamical information.

(3) Learn by 2D$-$CNN.

(4) Identify the critical point using \textit{confusion scheme}.


\noindent
In the following text we introduce in detail the structure and workflow of our
model and explain why it is powerful in the given task.

\subsection{Details of proposed method}

\textbf{How to represent network data as image-like structures?}

\begin{figure}
\centering
\epsfig{file=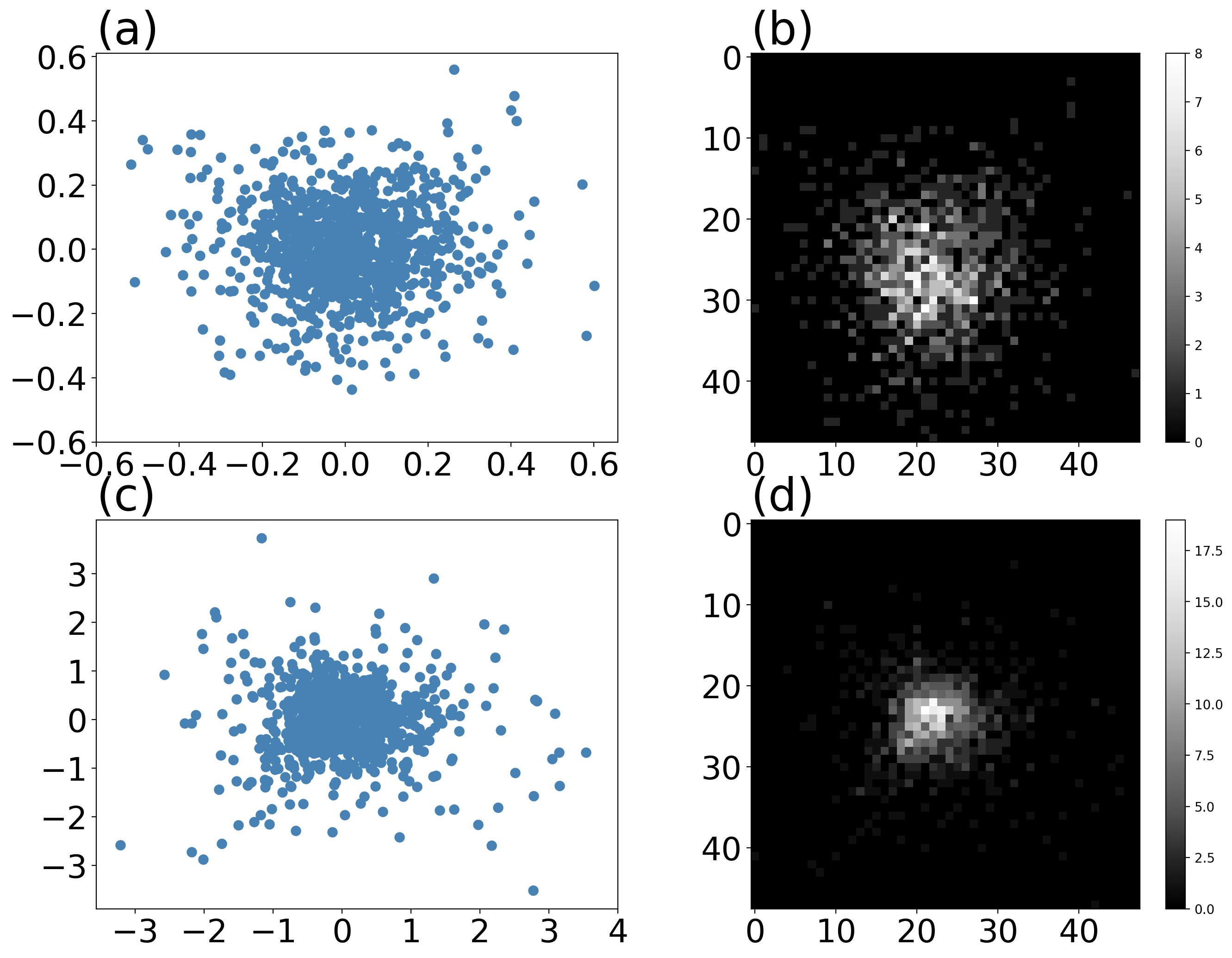,width=1\linewidth}
\caption{
\textit{Embedding network into two-dimensional space with deepwalk.}
(a, c) Embed an RR network with $N=1000$ and $\langle k\rangle=10$ (a) and
BA network with $N=1000$ and $m=m_0=3$ (c). Deepwalk is used to generate
the $d$-dimensional ($d\ll N$) representation for each node first. To reduce the
stochasticity of neural based algorithms, we use the PCA to extract two
principal components from the embedding space and greatly decrease the input
size to CNN. The horizontal and vertical axises represent the first and second
principal component, respectively.
(b, d) Discretize the embedding spaces and convert them into grayscale images.
To transform our embedded data into image-like structure, we split the space
into rectangle sub-region and color each region according to the number of
nodal representations in it (denoted by color bar). In this work, we create
images with the resolution of $48*48$, which means there're 48 sub-regions
in a row (column).
}
\label{fig:embedding}
\end{figure}

We aim to convert the high dimensional network data into 2D lattice, a
two dimensional Euclidean space discretized rectangles. A 2D lattice meets the
condition of \textit{spatial dependence}~\cite{tixier2017graph, tobler1970computer},
which indicates \textit{everything is related to everything else, but near things
are more related than distant things.} In a context of images, the closer the pixels
are, the more related they become. Consider a network denoted as $G(V,E)$ where $V$
and $E$ represent the node and edge sets, respectively. Classically, this
high-dimensional structure can be represented by its adjacency matrix $A$ or
Laplacian matrix $L$. Adjacency matrix is a square $(N,N)$ matrix where
$N$ is the size of $V$. It is symmetric in unweighted graphs and the $(i,j)^{th}$
entry $A_{ij}$ is the weight between two nodes $i$ and $j$ if there is an edge
$e_{ij}$ between them, or 0 otherwise. The Laplacian matrix can be obtained by
$L=D-A$, where $D$ is the diagonal degree matrix. The eigenvalue and eigenvector of
$L$ contains hidden information of the network. $L$ is largely used in
network representation learning algorithms based on matrix decomposition. To convert
this kind of data into an image, we do the following steps:

\textit{Graph embedding.} First, we use a graph representation learning algorithm to
embed the network data into a low dimensional underlying space.
The feature extracted from the network is the structural similarity between nodes, which
is indicated by the hyper distance of any two node representations.
For instance, neighbors tend to be close in the latent space compared to non-neighbors.
We get the representation by \textit{Deepwalk}~\cite{perozzi2014deepwalk}, which is a
graph embedding algorithm based on random-walk.
As many state-of-the-art network representation algorithms (e.g., Deepwalk) are neural, they are stochastic.
A given dimension will not be associated with the same latent space across networks or across several runs
on the same network. Therefore, PCA~\cite{Shlens:2014} is used to ensure these embeddings are comparable.
The PCA also serves an information maximization (compression)~\cite{tixier2017graph}
purpose as it greatly reduces the size of tensors fed to the CNN. In our work, PCA converts
the embedded data to 2-dimensional images which will be the input of CNN later.
Figure~\ref{fig:embedding} (a) shows the image which is obtained by embedding a random
regular (RR) network (with $N=1000$) into two-dimensional space.

\textit{Stacking node representations in 2D histograms.} Since the 2D representation
space is continuous and not able to be fed to CNN, we take a step forward to transform
the space into finite sized lattice, making it learnable for CNN. We first divide
the space into rectangle sub-regions and then color each sub-region by counting the
number of nodal representations falling into it. Accordingly, a grayscale image is
generated. As is represented in Fig.~\ref{fig:embedding} (b), the BA network is
converted into a grayscale image. The brighter a region is, the more nodes
there are. After discretization, the graph image turns to a stack of histograms representing
the number of nodes in each sub-area. It is worth noticing that the final
dimensionality does not depend on the size of the network (e.g., number of nodes
or edges). Networks with arbitrary size will be converted into images of the same
size.

\textit{Adding nodal dynamical information.} One of the key contributions of our
work is that we merge the nodal dynamical information into the 2D images. As there
are two states in the dynamics of SIS, we split our image into multiple channels,
with each channel represents one kind of nodal dynamical state information. For SIS,
we split it into two channels and there are only histograms of nodes in state $I$ ($S$)
in every single channel.
Fig.~\ref{fig:framework} (f) and (g) show the two-channel representations of
the original network with nodal dynamical states. In this manner, CNN will be able
to learn the structural and dynamical knowledge from the input data set, and make a
better decision than merely learn from part of them.

\textbf{How does CNN learn from images and what is the data to be learned?}

Deep learning on networks is a hot topic these years as graph neural networks
show the great performance on nodal semi-classification, link prediction, network
reconstruction, etc.~\cite{zhou2018graph,wu2019comprehensive,bronstein2017geometric,kipf2016semi}.
GCN is a simple but powerful model~\cite{kipf2016semi} which is able to learn from
network
structure and is widely used in a variety of network-related fields. However, the GCN is
unable to learn the dynamical information. What's worse,  the adjacency matrix is
needed to train the GCN and hence it is difficult to be applied in large-scale networks. We
take a very different way to learn the network's structural and dynamical
information at a time by converting the input network into
multi-channel, image-like 2D structures with CNN.
A CNN consists of three kinds of layers: the
convolution layer focuses on extracting information from its input by a continuous
sliding convolutional kernel; the pooling layer tries to reduce the size of data
which will
be feeded into next layer and to reduce the risk of over fitting; and the
fully-connected layer classifies the data into multiple ``groups''. In the context of
our work, there're two ``groups'' representing two different phases of the system.

In this work, the CNN contains four convolution layers, two max-pooling layers,
one fully-connected layer and one output layer. As is shown in fig~\ref{fig:framework}
(h), the sizes of the convolutional
filter (kernel) and pooling operator are $3*3$ and $2*2$, respectively. For the first
convolution-pooling block, 32 filters are employed in the convolution layer,
following with a $(2,2)$ max pooling layer which halves the input signal. The number
of filters in the subsequent convolution layer is increased to 64 to compensate
the loss of resolution. After the feature extraction process, the data enters a
128-neuron dense (fully connected) layer and the output layer will
give the probability that the input data is supposed to be in a specific phase.
Since our model is not sensitive to hyper-parameters, we decide these hyper-parameters,
together with some others, by a coarse grained grid search. We don't do a further
``fine tuning'' as the learning results have already been ``good''.
Dropout ~\cite{srivastava2014dropout} and $L2$ regularization are used to prevent
over fitting.

We use synchronous updating Monte Carlo simulation to obtain the training and test
data set for CNN. For the underlying neural network to learn the system dynamics,
proper inputs are needed. In an absorbing phase,
the nodal states are all identical (all $0$ in data set), rendering them improper
for training.
Near the phase transition point, i.e., when the value of $\lambda$ is in the
vicinity of $\lambda_c$ ($\left| \lambda-\lambda_c\right| \agt 0$), there
is a high probability for the system to be trapped in the absorbing state.
To overcome this, we use the quasi-stationary method~\cite{SCF:2016}
to prevent the system from entering the absorbing state. Specifically, whenever
the system tends to enter the absorbing state, we change its state to that of the
previous time step. This simple procedure allows us to obtain the legitimate
training and test sets for learning.

To make it more specific, consider the data set is basically ordered
along a tuning parameter $\lambda$, which is also the control parameter of the
phase transition. The data with a certain value of $\lambda$, will be described
by $\mathbf{s}(\lambda)$ and it consists of a set of grayscale images, each of
which represents the distribution of nodal 2D representations in the embedding
space. Nodal dynamical states are added to images by multi-channel. There are only
susceptible (infected) nodes in each channel. The brightness of each rectangle
sub-region indicates the amount of nodes in it.
The neural network is thought to be a high-level
mapping function that takes data $\mathbf{s}(\lambda)$ to infer the probability
distribution $\mathcal{F}(\mathbf{s}(\lambda))=p_0$, where $p_0$ ($1-p_0$)
represents the inferred probability that $\mathbf{s}(\lambda)$ is in the phase of
absorbing (active) state. This can be considered directly with the probability
$L(\lambda)=H(\lambda-\lambda_c)$, where $\lambda_c$ is the ground truth
(fake truth) that is preset in advance and $H$ represents the unit step function
$H(x)=\frac{d}{dx}max\{x,0\},x\neq 0$.
In the training procedure, a large data set of different
$\mathbf{s}(\lambda)$ and corresponding labels $L(\lambda)$ are fed to the neural
network and this network makes
$\mathcal{F}(\mathbf{s}(\lambda))\rightarrow L(\lambda)$ by optimizing its parameter
set $\mathbf{P}_\mathcal{F}$ to minimize a cost function
$C(\mathcal{F}(\mathbf{s}(\lambda)),L(\lambda))$. The cost function
evaluates the gap between the network's judgment and the real answer
quantitatively. It can be minimized through a lot of optimization method such as
stochastic gradient descent or $ADAM$~\cite{kingma2014adam}.
In this work, input data are a set of $(48,48,2)$ images
($48$ for resolution and $2$ for number of channel) and the
corresponding labels are a vector of $0$ and $1$, representing absorbing and
active phase, respectively.

\begin{figure}
\centering
\epsfig{file=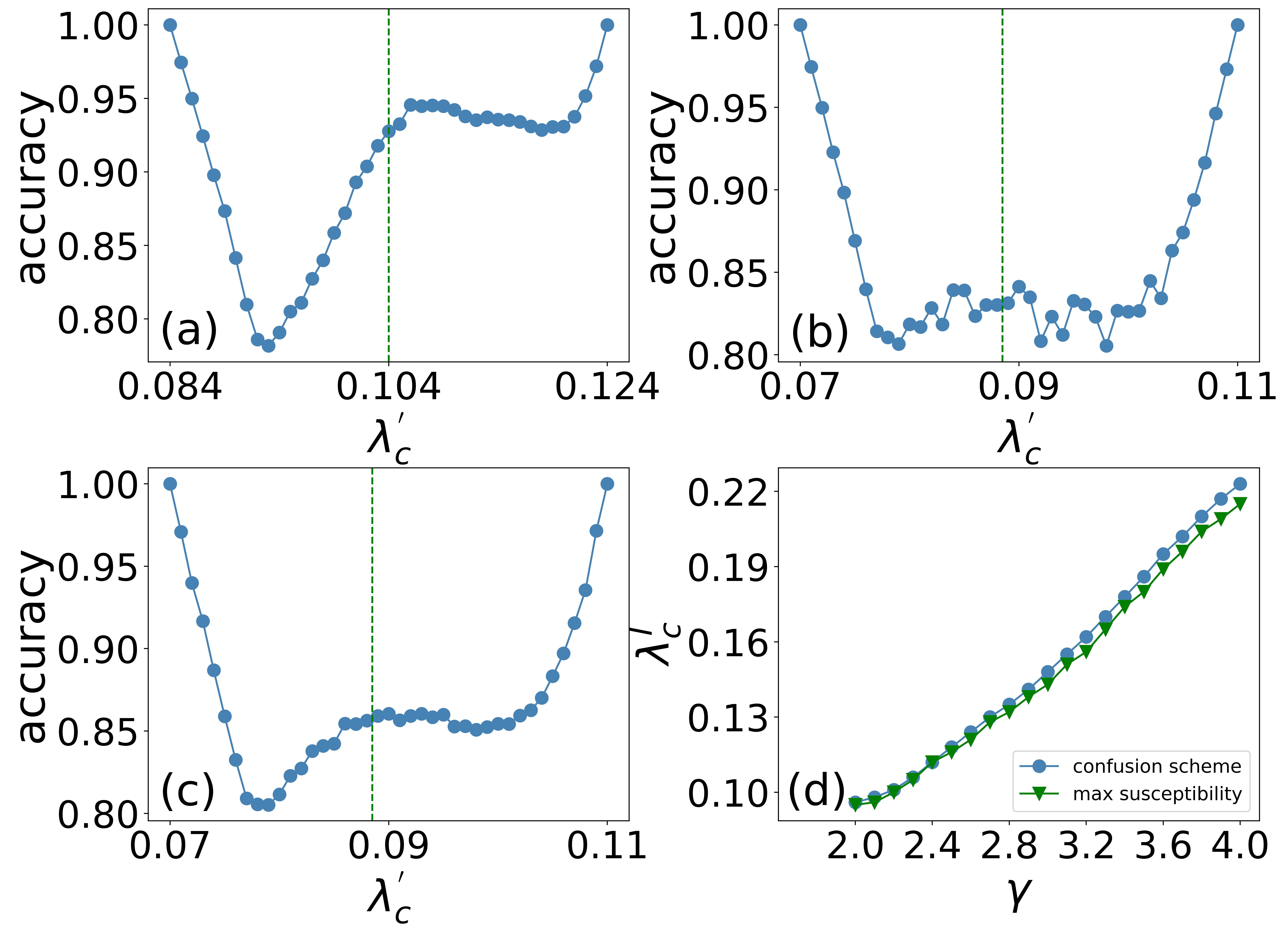,width=1\linewidth}
\caption{
\textit{Learning phase transition on homogeneous and heterogeneous networks with
deepwalk.}
(a) For a random regular network of size $N=1000$ and average degree
$\langle k\rangle=10$, over model generates an overall $W$-shape curve of the
classification accuracy versus the assumed threshold value. The peak of
susceptibility is at $0.104$, as indicated by the green vertical line (the same
legend holds below).
(b) For a $BA$ network of size $N=1000$ and $m=m_0=3$, the neural network fails to
yield a $W$-shape accuracy curve: the accuracy has a $U$-shape.
(c) For the same $BA$ network as (b), the accuracy curve obtained by incorporating
$Poincar\acute{e}$ embedding into the training process. In this case, the accuracy
curve exhibits a $W$-shape, rendering detectable the phase transition. The transition
point can also be identified accurately.
(d) For scale-free networks generated by the
\textit{uncorrelated configuration model} (UCM) of size
$N=1000$, max degree $k_{max}=\sqrt{N}\approx32$ and the power law exponent $\gamma$ in the range $[2.0,4.0]$,
the epidemical thresholds obtained by susceptibility measure and our model are shown.
Blue circles and green triangles represent the results given by susceptibility and
our model, respectively.
}
\label{fig:deepwalk}
\end{figure}

\textbf{How are the critical point inferred by \textit{confusion scheme}?}

As is demonstrated in the previous sub-section, it is essential to have a correct set
of labels to make the CNN draw an unbiased conclusion. However, in some cases
the label, which is supposed to be a kind of prior knowledge, cannot be easily
obtained or guaranteed flawless. These dirty labels will significantly decrease the
performance of the model~\cite{mypaper1}.
We use the \textit{confusion scheme} that can make precise prediction
without any prior knowledge about the labels~\cite{vNLH:2017}. The only difference
between confusion scheme and original supervised learning is that the labels
of confusion scheme method are man-made guesses instead of ground truths. As
our data are indexed by order parameter $\lambda$, we can avoid using any prior
knowledge by making a ``good guess'' of the true transition point. The number
of candidates in total is $\mathcal{N}+1$ where $\mathcal{N}$ is the number
of different values of $\lambda$ existing in our data set.
It's completely affordable to
perform such a ``brute force'' guess and the result can be inferred
by our neural network.

Figure~\ref{fig:framework} (I) shows how the \textit{confusion scheme} work in detail.
Suppose we have a set of unlabeled network configuration data ranging from
$\lambda_{min}$ to $\lambda_{max}$. Let $\lambda_c$ be the unknown true
threshold value, where $\lambda_{min}\leq\lambda_c\leq\lambda_{max}$. We
assign tentative labels to the data set by assuming that the threshold
is $\lambda_c^{\prime}$, where $\lambda_{min}\leq\lambda_c^{\prime}\leq\lambda_{max}$.
We assign label zero to all configurations for
$\lambda\leq\lambda_c^{\prime}$ and label one to those with $\lambda>\lambda_c^{\prime}$.
We then choose a number of closely located values of $\lambda_c^{\prime}$. For
each value of $\lambda_c^{\prime}$, we conduct the training and obtain the
classification accuracy. Ideally, we expect the accuracy to exhibit a
$W$-shape curve versus $\lambda_c^{\prime}$. For
${\lambda_c}^{\prime}=\lambda_{min}$, every row of the training set is labeled as
one, so the neural network regards the data of every pattern as in the
activation phase, giving rise to $100\%$ accuracy of prediction. Similar
result is expected for $\lambda_c^{\prime}=\lambda_{max}$. For
${\lambda_c}^{\prime}=\lambda_c$, the method reduces to supervised learning
because the tentative or ``fake'' labels happen to be correct under the
circumstance. The neural network
can yield a high classification accuracy in this case~\cite{mypaper1}.
For $\lambda_{min}<{\lambda_c}^{\prime}<\lambda_c$ or
$\lambda_{c}<{\lambda_c}^{\prime}<\lambda_{max}$, the neural network will be
``confused'' for some data whose labels are exactly opposite to the true
values, thereby leading to a decrease in the accuracy. Overall, a $W$-shape
of the dependence of the accuracy on the value of $\lambda_c^{\prime}$ occurs,
where the location of the peak in the middle corresponds to the identified
threshold. If there is no phase transition in the threshold range
$[\lambda_{min},\lambda_{max}]$, the accuracy versus $\lambda_c^{\prime}$ would
exhibit a $U$ shape. The emergence of a $W$-shape curve is thus
an indication that there is phase transition in the system and the
correct transition point (or threshold) can be identified accordingly without
requiring any prior knowledge about the labels.

\section{The necessity of hyperbolic embedding} \label{sec:hyperbolic}

To gain insight, we test our machine learning framework in both homogeneous
and heterogeneous complex networks in Fig.~\ref{fig:deepwalk}.
Figure~\ref{fig:deepwalk} (a) shows that our method
can successfully identify the threshold in random regular networks with mean degree
$\langle k\rangle=10$. Numerical simulation of the SIS dynamics reveals a
second-order phase transition at $\lambda\approx0.1040$ while the confusion scheme
yields an overall $W$-shape behavior and the position of the middle peak
occurs at $\lambda_c^\prime\approx0.1050$, which is
indistinguishable from the actual threshold value. In general, the proposed framework
is quite effective for detecting phase transition with an accurate identification of
the epidemical threshold for homogeneous networks.

For heterogeneous networks such as scale-free and $BA$ networks, this learning model
is not always effective. As demonstrated in Fig.~\ref{fig:deepwalk} (b), the accuracy
curve does not exhibit an apparent $W$ shape in the chosen threshold range, whereas
an actual phase transition occurs at $0.0893$. A possible reason is
that, in a typical heterogeneous network, the majority of nodes turn out to gather
in a small area of the central part of the embedding space, as illustrated in
Figs.~\ref{fig:embedding} (c) and (d). This pattern may lead to the decrease of
information and thus the learning performance deteriorates.
It is known that the random walk based algorithm like deepwalk evaluates the
similarity between two nodes by the probability that one node exists in the
random walk sequence of the other node. For scale-free networks like $BA$ model,
there are
often nodes of small degree in the neighborhood of a hub. In this way, it is
possible that the deepwalk considers these two types of nodes similar during
multiple round of training. It leads to the existence of the previously
mentioned pattern.
One possible solution is
to increase the resolution of the image. We've tested the scenario where images are
sized
$(128,128,2)$ instead of $(48,48,2)$ and the $W$ shape can be obtained. However,
because the time complexity against the resolution of images is quadratic,
the computational cost increases sharply as the image resolution grows, implying
it is impossible to be used in large scale networks.

\begin{figure}
\centering
\epsfig{file=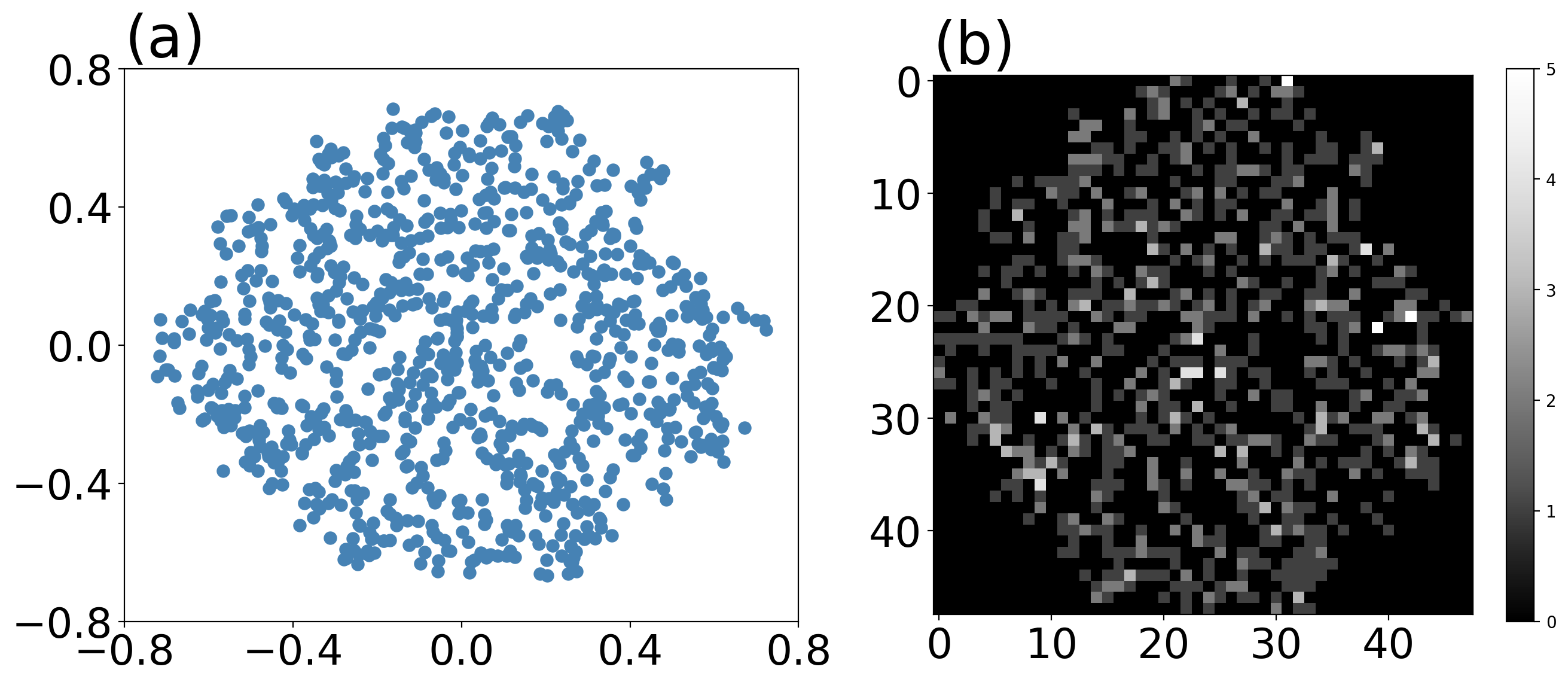,width=1\linewidth}
\caption{
\textit{Embedding network into two-dimensional space with
$Poincar\acute{e}$ embedding.}
(a) Embed BA network with $N=1000$ and $m=m_0=3$ into a two dimensional hyperbolic
disk. Each blue node in the figure is a two dimensional vector representing
a vertex in the original network in the embedded space.
(b) Discretize the embedding spaces and convert them into grayscale images.
To transform our embedded data into image-like structure, we split the space
into rectangle sub-region and color each region according to the number of
nodal representations in it (denoted by color bar). In this work, we create
images with the resolution of $48\times48$, which means there're 48 sub-regions
in a row (column).
}
\label{fig:Poincare}
\end{figure}

Hence, a better embedding algorithm is needed to make the image look less "crowded".
Next we pay attention to hyperbolic embedding. In mathematics, a hyperbolic
space is a homogeneous space that has a constant negative curvature
~\cite{bonahon2009low}. Hyperbolic space is suitable to model hierarchical data and
this pattern is existing in complex networks widely, especially in real-world
networks. Many previous works represent networks in hyperbolic disk
~\cite{muscoloni2017machine,papadopoulos2012popularity}, where the relationship
between Euclidean and hyperbolic distance $r_e$ and $r_h$ is $r_e=tanhr_h/2$. It
means that for a fixed Euclidean distance, hyperbolic distance in the peripheral area
is much ``longer'' than it is in the central area of the disk.
Here we exploit $Poincar\acute{e}$ embedding
which embeds the high dimensional networks into a low dimensional $Poincar\acute{e}$
ball model~\cite{nickel2017poincare}. Figure~\ref{fig:Poincare}
represents the image-like structure obtained by $Poincar\acute{e}$
embedding. As is shown, the nodes in the embedded space distributed equally in the
embedding space instead of gathering in the central area, reducing the information
loss compared to those with deepwalk algorithm [Figs.~\ref{fig:embedding} (c) and (d)].

\begin{figure}
\centering
\epsfig{file=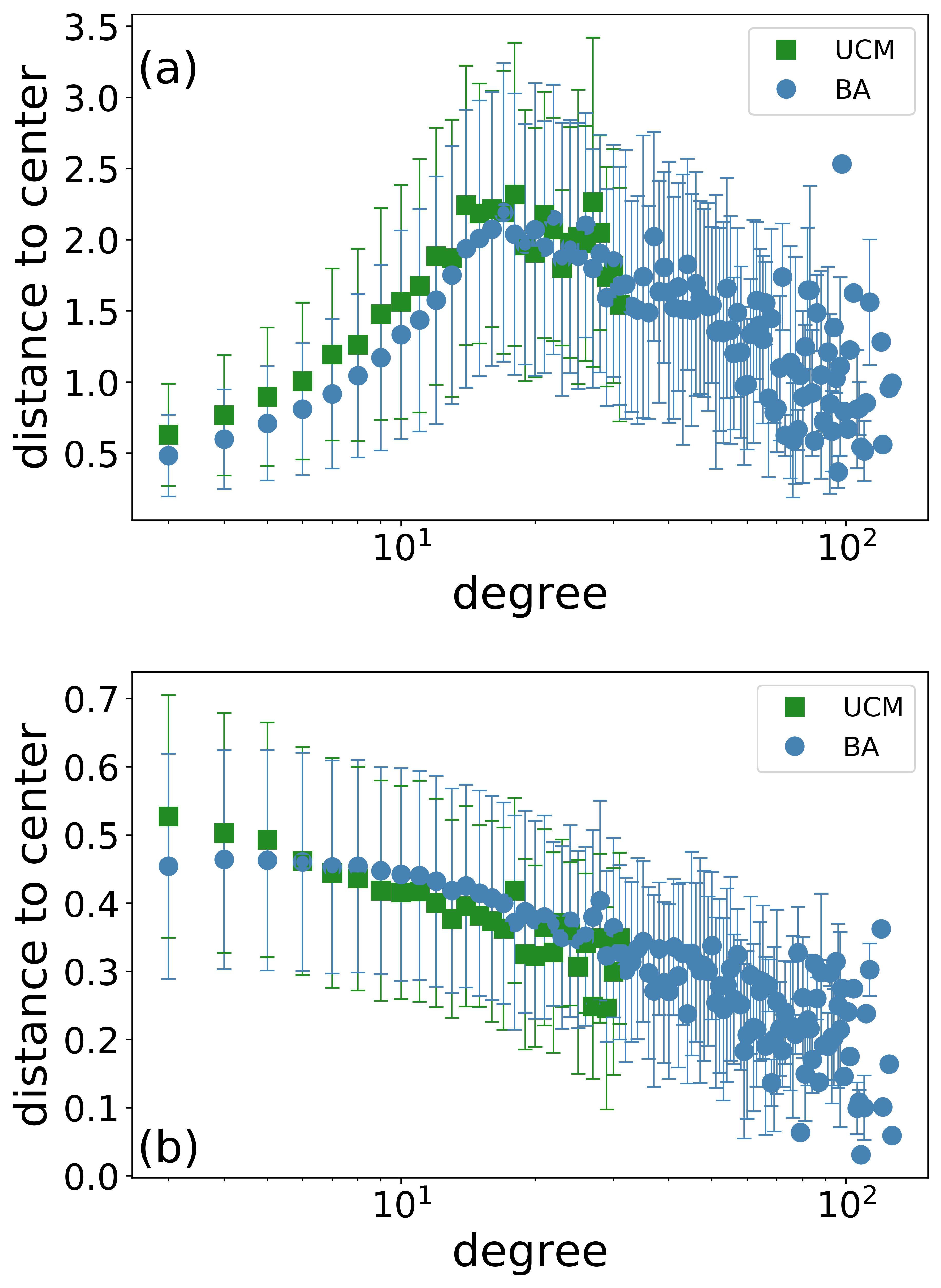,width=1\linewidth}
\caption{
\textit{Nodal distribution in the 2D embedding space} of (a) \textit{Deepwalk} and (b)
$Poincar\acute{e}$ embedding for $BA$ networks of size $N=1000$ and $m=m_0=3$
(blue circles) and scale-free networks with $\gamma=2.5$ and $N=1000$ generated by UCM
(green squares).
Axises $x$ and $y$ represent the degree of nodes and
the average Euclidean distance from nodes with a specific degree to the center of
the space.
}
\label{fig:distribution}
\end{figure}

Figure~\ref{fig:distribution} describes another reason why the $Poincar\acute{e}$
embedding outperforms the random walk based method. It is known that
in the SIS spreading process, the hub nodes are the main spreaders.
Nodes with large degree are considered more important than those with small
degree~\cite{CP:2012}, and the latter have little chance to be infected multiple times which accordingly
provides less dynamical information than hubs. In an ideal scenario, we want to maximize the
information given by nodes with large degree thus making the learning easier for
CNN. Figure~\ref{fig:distribution} (a) and (b) represent the relationship between
nodal degree and average Euclidean distance of nodes to the center of the 2D embedding
space. For \textit{Deepwalk}, both the nodes with small and large degrees tend to
locate in the
central area of the space, which deteriorates the influence of hubs. In the
discretized images with dynamical states, the central area and
the periphery will both be dim. However, for the $Poincar\acute{e}$ embedding, hubs and
leaves are distributed in the center and periphery, respectively. This pattern
strengthens the influence of large-degree nodes: first, nodes do not gather in a
very small area which reduces the information loss; second, hubs are in the central
area and a bright spot is formed. In this way CNN will
learn more dynamical information and accordingly get a better performance.

As represented in Figs.~\ref{fig:deepwalk} (c) and (d), we test
$Poincar\acute{e}$ embedding
in BA and scale-free networks. In Fig.~\ref{fig:deepwalk} (c),
for the same BA network as in Fig~\ref{fig:deepwalk} (b),
the \textit{confusion scheme} can successfully detect the phase transition
and identify the critical point. In Fig.~\ref{fig:deepwalk} (d),
for scale-free networks generated by
\textit{uncorrelated configuration model} (UCM) with size $N=1000$ and
the power law exponent $\gamma$ in the range $\gamma\in[2.0,4.0]$,
with the help of $Poincar\acute{e}$ embedding, our
machine learning framework does a great job in identifying the epidemic threshold
accurately.

\section{Application to real world networks} \label{sec:real}

\begin{figure}
\centering
\epsfig{file=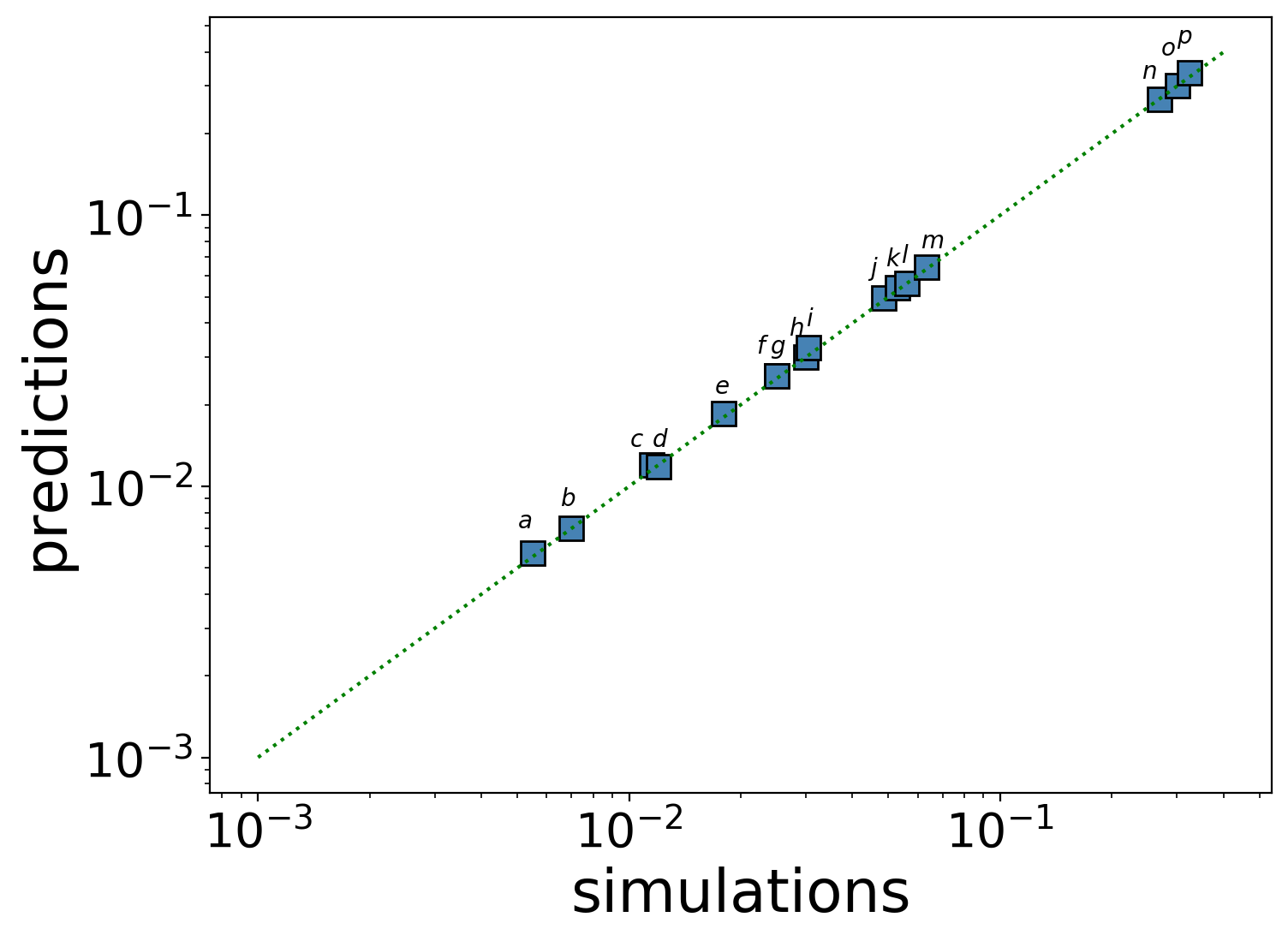,width=1\linewidth}
\caption{
\textit{Performance on more real-world networks.}
Identified versus simulated epidemic threshold with $16$ real-world network data set.
$a$ to $p$ represent \textit{Reactome}, \textit{Facebook}, \textit{Astro-Ph},
\textit{Brightkite}, \textit{Caida}, \textit{Hamsterster-full},
\textit{Hamster Friendships}, \textit{Jazz Musicians}, \textit{Pretty Good Privacy},
\textit{Gnutella5}, \textit{Gnutella6}, \textit{Arenas-email}, \textit{Gnutella4},
\textit{US Power Grid}, \textit{RV}, \textit{Euroroad}, respectively. Detail descriptions
of the networks we use can be found in appendix.
}
\label{fig:16real}
\end{figure}

We test the performance of our model in real-world networks. Figure~\ref{fig:tworeal}
illustrates the results for two real-world datasets. Although \textit{Deepwalk} can get
the final results sometimes,
we use $Poincar\acute{e}$ embedding
as it is well suited for processing hierarchical topology, which is a common
feature in real-world complex networks. Results in more real-world networks
are shown in Fig.~\ref{fig:16real}. It is proved that our model gives a very accurate
identified epidemical thresholds in almost every network.
Details of the real-world data set we use and the corresponding results
are listed in the appendix:

\begin{figure}
\centering
\epsfig{file=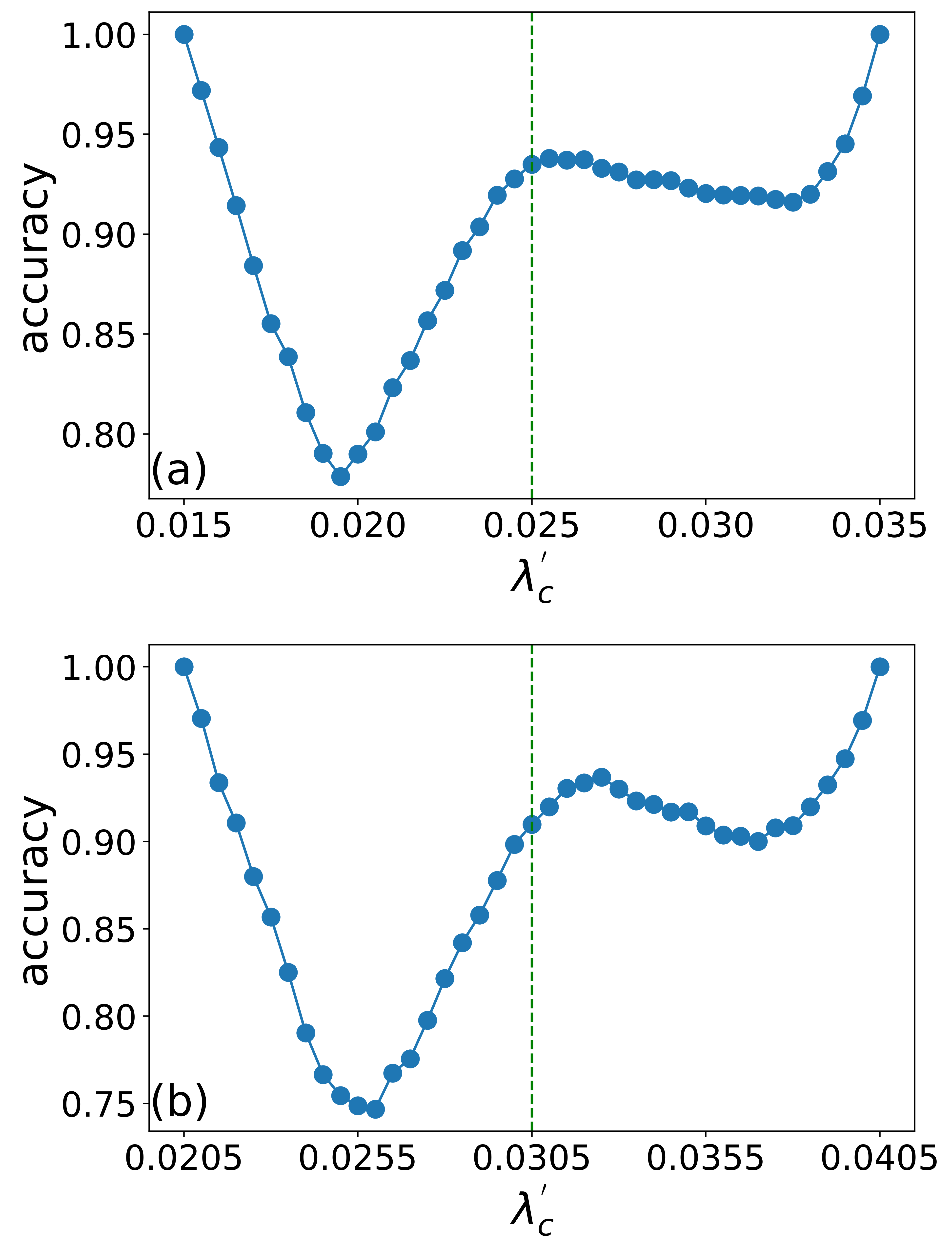,width=1\linewidth}
\caption{
\textit{Performance in two real-world networks.}
(a) \textit{Petster Friendship Hamster} data set.
(b) \textit{Pretty Good Privacy} data set.
For the details of these empirical networks and the corresponding results, please see
the appendix.
}
\label{fig:tworeal}
\end{figure}

\section{Conclusion and discussion} \label{sec:conclusion}

In this paper, we propose a machine learning framework to identify the threshold
of epidemic dynamics in complex networks and find that this method performs
pretty well on a variety of synthetic and empirical network topologies. The framework
is end-to-end, computationally efficient, robust and universally applicable.
We first convert the high dimensional networks into 2D images, in which the graph
representation learning algorithms play an important role. While random walk
based \textit{Deepwalk} is simple and fast in computation, it fails to generate good
pattern for CNN in some heterogeneous networks. Hence, the hyperbolic embedding
method is introduced and applied on a large amount of network topologies. With
its powerful ability in maximizing the information of hubs and representing
networks with a hierarchical structure, $Poincar\acute{e}$ embedding is exploited
in the majority of cases. We creatively add nodal dynamical states into the
image with multiple channels: each channel represents the sub-network
consisting of
nodes in the state of $S$ ($I$). In this way, the following CNN is capable of
learning dynamical and structural information of the network at a time. The CNN
outputs a classification probability for input images with each $\lambda$ and
finally the \textit{confusion scheme} automatically detects if there's a
phase transition in the input data and determines the epidemical threshold without any
prior knowledge. Compared to our previous work, the proposed framework learns the topological
and dynamical information simultaneously to give a better identification of the transition point.
It is free from any kind of feature engineering,
included the ``sampling methods'' we use in the previous work.

There still remains a lot of questions where further researches are needed. For
instance, how to scale the input network up to a super large size while limiting
the computational cost in a reasonable range; how to take a step forward to
make the model fit different network topologies after a single training; how
to apply our method in time-varying and multiplex networks, etc.

It is proved by plenty of experiments in $17$ empirical data set that our machine
learning framework performs great in almost every case. Our main contributions are
two-fold. First, the proposition of our framework opened a new direction to learn
and identify phase transition point of dynamics in complex networks, using machine
learning techniques. Second, our work combines both topological and dynamical
information of a network, exploring a new way to solve the threshold-related issues.

\acknowledgements

This work was supported by the National Natural Science Foundation of China
under Grant Nos.~11575041, 61802321, 11872182 and 11835003, the Natural Science Foundation
of Shanghai under Grant No.~18ZR1412200, and the Science and Technology Commission
of Shanghai Municipality under Grant No. 18dz2271000.

\section{Appendix} \label{appendix}

The detailed information of the empirical networks:

\textit{CAIDA}~\cite{LKF:2007}. This is the undirected network of autonomous
systems of the Internet from the CAIDA project, collected in 2007. Nodes
are autonomous systems (AS) and edges represent communication.

\textit{Brightkite}~\cite{CML:2011}. This undirected network contains user-user
friendship relations from Brightkite, a former location-based social network
were users share their locations. A node represents a user and an edge
indicates that a friendship exists between a pair of users.

\textit{Astro$-$Ph}~\cite{LKF:2007}. This is the collaboration network of
authors of scientific papers from the arXiv's Astrophysics (astro-ph)
section. An edge between two authors represents a joint publication.

\textit{PGP (Pretty Good Privacy)}~\cite{BPSDA:2004}. This is the interaction
network of users of the Pretty Good Privacy (PGP) protocol. The network
has only one giant connected component.

\textit{RV (Route Views)}~\cite{LKF:2007}. This is an undirected network of
the autonomous system of the Internet.

\textit{Facebook}~\cite{LM:2012}. This data set consists of ``circles'' (or
``friend lists'') from Facebook. Facebook data were collected from survey
participants using this Facebook app. The data set includes nodal features
(profiles), circles, and ego networks.

\textit{Gnutella}~\cite{RFI:2002}. This is a sequence of snapshots of the
Gnutella peer-to-peer file sharing network from August 2002. There are
altogether nine snapshots of Gnutella network collected in August 2002.
Nodes represent hosts in the Gnutella network and edges are connections
between the hosts.

\textit{Reactome}~\cite{joshi2005reactome} This is a network of protein–protein
interactions in the species Homo sapiens, i.e., in Humans. The data is curated
by the Reactome project, an open online database of biological pathways.
Nodes represent protein and edges represent interactions.

\textit{Hamsterster-full}~\cite{hamsterster} This Network contains friendships
and familylinks between users of the website hamsterster.com. Nodes and edges
represent users and user-to-user relationship, respectively.

\textit{Hamsterster Friendships}~\cite{hamster} This Network contains friendships
and familylinks between users of the website hamsterster.com. Nodes and edges
represent users and user-to-user relationship, respectively.

\textit{Jazz Musicians}~\cite{gleiser2003community} This is the collaboration network
between Jazz musicians. Each node is a Jazz musician and an edge denotes that two
musicians have played together in a band. The data was collected in 2003. Nodes
represent musicians and edges represent collaborations.

\textit{Arenas-email}~\cite{guimera2003self} This is the email communication network
at the University Rovira i Virgili in Tarragona in the south of Catalonia in Spain.
Nodes are users and each edge represents that at least one email was sent. The direction
of emails or the number of emails are not stored.

\textit{US power grid}~\cite{watts1998collective} This undirected network contains
information about the power grid of the Western States of the United States of America.
An edge represents a power supply line. A node is either a generator, a transformator or
a substation.

\textit{Euroroad}~\cite{vsubelj2011robust} This is the international E-road network, a
road network located mostly in Europe. The network is undirected; nodes represent cities
and an edge between two nodes denotes that they are connected by an E-road.


\bibliography{CNN}

\end{document}